\documentclass[usenatbib]{mn2e}
\usepackage{natbib}
\usepackage{epsf}
\usepackage{graphicx}

\title[Optical/IR jets in neutron star X-ray binaries]{Evidence for a jet contribution to the optical/infrared light of neutron star X-ray binaries}
\author[D. M. Russell, R. P. Fender \& P. G. Jonker]
{D. M. Russell$^{1}$\thanks{Email: davidr@phys.soton.ac.uk},
R. P. Fender$^{1}$, P. G. Jonker$^{2,3,4}$
\\
$^1$School of Physics \& Astronomy, University of Southampton, Highfield, Southampton, SO17 1BJ, UK\\
$^2$SRON Netherlands Institute for Space Research, Sorbonnelaan 2, 3584 CA Utrecht, Netherlands\\
$^3$Harvard-Smithsonian Center for Astrophysics, 60 Garden Street, MS 83, Cambridge, MA 02138, USA\\
$^4$Astronomical Institute, Utrecht University, P.O.Box 80000, 3508 TA, Utrecht, The Netherlands\\
}

\begin{document}
\maketitle

\begin{abstract}
Optical/near-infrared (optical/NIR; OIR) light from low-mass neutron star X-ray binaries (NSXBs) in outburst is traditionally thought to be thermal emission from the accretion disc. Here we present a comprehensive collection of quasi-simultaneous OIR and X-ray data from 19 low-magnetic field NSXBs, including new observations of three sources: 4U 0614+09, LMC X--2 and GX 349+2. The average radio--OIR spectrum for NSXBs is $\alpha\approx +0.2$ (where $L_{\nu} \propto \nu^{\alpha}$) at least at high luminosities when the radio jet is detected. This is comparable to, but slightly more inverted than the $\alpha\approx 0.0$ found for black hole X-ray binaries. The OIR spectra and relations between OIR and X-ray fluxes are compared to those expected if the OIR emission is dominated by thermal emission from an X-ray or viscously heated disc, or synchrotron emission from the inner regions of the jets. We find that thermal emission due to X-ray reprocessing can explain all the data except at high luminosities for some NSXBs, namely the atolls and millisecond X-ray pulsars (MSXPs). Optically thin synchrotron emission from the jets (with an observed OIR spectral index of $\alpha_{\rm thin}< 0$) dominate the NIR light above $L_{\rm X} \approx 10^{36}$ erg s$^{-1}$ and the optical above $L_{\rm X} \approx 10^{37}$ erg s$^{-1}$ in these systems. For NSXB Z-sources, the OIR observations can be explained by X-ray reprocessing alone, although synchrotron emission may make a low level contribution to the NIR, and could dominate the OIR in one or two cases.

\end{abstract}


\section{Introduction}

In low-mass neutron star X-ray binary (NSXB) systems, matter is accreted onto a neutron star from a companion via a disc which radiates in the optical, ultraviolet and X-ray regimes.  Some of the accreting matter and energy in X-ray binaries (XBs) can be released from the system through collimated outflows (jets).  Evidence for jets associated with NSXBs date back more than a decade \citep*{stewet93,bradet99,fomaet01} but not until recently has the evidence emerged from any waveband other than the radio. \cite*{callet02} found that infrared $K$-band flaring in the `Z-source' NSXB GX 17+2 could not be explained by an X-ray driven wind or reprocessed X-rays, but shared many properties with the radio (i.e. jet) variability previously seen in the same source. It is worth noting that the near-infrared (NIR) counterpart of GX 13+1 is also largely variable \citep{charna92}. Recently, \cite*{miglet06} for the first time spectrally detected optically thin synchrotron emission from the jets of an `atoll' NSXB 4U 0614+09 in the mid-infrared.

In the optical regime, many spectral and timing studies of NSXBs have established the presence of an accretion disc reprocessing X-ray photons to optical wavelengths \citep*[e.g.][]{mcclet79,lawret83,konget00,mcgoet03,hyneet06}. In quiescence, the companion star can come to dominate the optical/NIR (OIR) emission \citep*[e.g.][]{thoret78,chevet89,shahet93}, as is the case in high-mass X-ray binaries \citep*[e.g.][]{vandet72} and low-mass X-ray binaries that accrete from a giant star. X-ray reprocessing is generally thought to dominate the OIR emission of non-quiescent NSXBs (see \citealt*{vanpet95} for a review; see also \citealt*{chenet97,charco06}). In low-mass black hole candidate X-ray binaries (BHXBs), additional OIR emission mechanisms have been observed, in particular from the viscously heated disc \citep*[e.g.][]{kuul98,brocet01,homaet05} and from compact jets (e.g. \citealt*{hanet92,fend01,corbet02,hyneet03,brocet04,buxtet04}; Russell et al. 2006).

Recently, an anomalous transient NIR excess has been observed in a number of millisecond X-ray pulsar (MSXP) NSXBs at high luminosities, which is equivocal in nature. The source most studied is SAX J1808.4--3658, for which \cite*{wanget01} found a NIR flux almost one order of magnitude too bright to originate from X-ray heating. The NIR flux density was comparable to a radio detection of 0.8 mJy (with a flat 2.5--8.6 GHz spectrum) seen one week after the NIR excess. \cite*{greeet06} also reported an $I$-band excess in a different outburst of the same source, which they attributed to synchrotron emission.  In addition, a variable $I$- and $R$-band excess in XTE J0929--314 seen on the same day as a radio detection \citep*{gileet05}, and a transient NIR excess in XTE J1814--338 \citep*{krauet05} and IGR J00291+5934 \citep{torret07} were all interpreted as synchrotron emission from the jets in the systems. The NIR excess appears to be ubiquitously absent at lower luminosities.

Steady, partially self-absorbed jets probably exist in low-magnetic field ($B \la 10^{\sim 11}$ G) NSXBs in hard X-ray states \citep*{miglfe06,mass06}. These include `atoll-type' sources in the `island' state, `Z-type' sources in the `horizontal branch' and possibly the `normal branch'\footnote{These states correspond to the position of the source in the X-ray colour--colour diagram; see \citealt*{hasiva89}.} and transients at low accretion rates ($L_{\rm X}\la 0.1 L_{\rm Edd}$) such as MSXPs.  \cite{mass06} shows on theoretical grounds that the existence of jets in NSXBs depends on the magnetic field and mass accretion rate, and the conditions required for jet ejection are probably fulfilled for most of the NSXBs (Z-sources, atolls and MSXPs) with known magnetic field strengths. In Russell et al. (2006; hereafter \citealt{paper1}), we predicted that emission from the jets could contribute or even dominate the OIR light of NSXBs at high X-ray luminosities, for sources in a hard X-ray state. This is expected if $L_{\rm OPT,jet}\propto L_{\rm X}^{1.4}$ \citep[a theoretical relation supported by radio--X-ray observations;][]{miglfe06} and $L_{\rm OPT,XR}\propto L_{\rm X}^{0.5}$ (where XR is X-ray reprocessing in the disc), as is predicted and observed (\citealt*{vanpet94,paper1}). In the sample of \citealt{paper1} there were too little data at high X-ray luminosities to test this prediction for NSXBs.

In this paper we analyse quasi-simultaneous OIR and X-ray data and OIR spectra of a large, comprehensive sample of NSXBs in order to constrain the $L_{\rm X}$-dependent dominating OIR emission processes and to test the above prediction. Data are collected from transient NSXBs: atolls, MSXPs and Z-sources. We use the technique \citep{paper1} of comparing the relations between the quasi-simultaneous OIR and X-ray fluxes with the relations predicted from models of three OIR emission processes: a viscously heated disc, an X-ray reprocessing disc and synchrotron-emitting jets. We also inspect the spectral index of the OIR continuum, which differs largely between emission from the disc and from the jets. If the radio/OIR jet behaviour is ubiquitous in NSXBs, its properties will constrain many parameters in these systems, in particular the power of the jets (if the jet radiative efficiency can be constrained) and the wavelength-dependent jet contribution at a given X-ray luminosity.

\section{methodology \& results}

A wealth of OIR and X-ray data from atolls, Z-sources and MSXPs were collected from the literature and converted to intrinsic luminosities (or monochromatic luminosities) in the same manner as described in Section 2 of \citealt{paper1}. Much of the data we use here were obtained for \citealt{paper1}, however we have sought additional quasi-simultaneous OIR and X-ray data \citep[and OIR spectral energy distributions; SEDs, which were not collected for][]{paper1} for the purposes of obtaining a comprehensive sample for this work (see Tables 1 and 2 for the references of the new data).

Estimates of the distance to each source, the optical extinction $A_{\rm V}$ and the HI absorption column $N_{\rm H}$ used here are listed in Table 2 of \citealt{paper1} and Table 1 of this paper. No data were included for which the OIR contribution of the companion star could be significant (i.e. quiescence for most sources) unless this contribution is known. For these data we subtracted the star light contribution. For the Z-sources, OIR data were only included when the OIR fluxes were significantly higher than the lowest measured for each source in each waveband, to ensure minimal contamination from the companion. When two or more OIR wavebands were quasi-simultaneous, OIR SEDs were produced in order to spectrally constrain the origin of the OIR emission. In addition to the data collected from the literature, we obtained OIR photometry of three NSXBs using two telescopes; the observations and data reduction are discussed in the following subsections.

While quasi-simultaneity was achieved in most cases with use of the \emph{RXTE} ASM X-ray daily monitoring, this is not possible for some historical OIR observations. In these cases we include the data only in the OIR SEDs. Similarly, for some sources OIR--X-ray quasi-simultaneity was achieved, but there was only one OIR waveband available so OIR colours and SEDs were not obtained (e.g. GX 17+2). Quasi-simultaneous OIR--X-ray data were available for five sources (LMC X--2, Cyg X--2, Cen X--4, 4U 0614+09 and Aql X--1) before the advent of the \emph{RXTE} \citep*{kaluet80,caniet80,charet80,bonnet89,hasiet90,vanpet90,machet90}.

\begin{table*}
\begin{center}
\caption{Properties and data collected for the 11 NSXBs not included in \citealt{paper1} and IGR J00291+5934, whose parameters have been updated since Paper I (see Table 2 of \citealt{paper1} for the properties of the remaining NSXBs).}
\begin{tabular}{lllllllllll}
\hline
Source &Type&Distance&Period &$M_{\rm co}$ &$M_{\rm cs}$&$A_{\rm V}$&$N_{\rm H}$ / &$q_{\rm cs}$&$\Delta t$ /&Fluxes - \\
= alternative &&/ kpc &/ hours&/ $M_\odot$&/ $M_\odot$&&10$^{21} cm^{-2}$&(band,&days &data \\
name &&(ref) &(ref) &(ref) &(ref)&(ref) &(ref) &ref)& &refs\\
(I)&(II)&(III)&(IV)&(V)&(VI)&(VII)&(VIII)&(IX)&(X)&(XI)\\
\hline
IGR&MSXP&2.8$\pm$1.0&2.457&1.4&0.039--&2.5$\pm$0.3&4.64$\pm$0.58&-&1.0&1, 36\\
\vspace{2mm}
J00291+5934&&(1)&(13)&(16)&0.160 (16)&(1)    &(1)         &&&\\
LMC X--2&Z-source&50$\pm$10&8.16&$\sim 1.4$&$\sim 1.2$&0.15&0.91$\pm$0.07&-&0.5&20, 37,\\
\vspace{2mm}
= 4U 0520--72&&(2, 3)&(14)&(17)&(17)&(20)    &(29)         &&&38\\
XTE J0929--314&MSXP&$>5$ (4)       &0.726&-&$\sim 0.008$&0.42$\pm$0.10&0.76$\pm$0.24&-&1.0&21, 37\\
\vspace{2mm}
= INTREF 390  &&($\sim 8\pm 3$)&(14)&&(4)&(21)          &(30)          &&&\\
Cir X--1&Z-source&9.2$\pm$1.4&398&-&4$\pm$1&10.5$\pm$1.5&19$\pm$3&-&0.2&37, 39,\\
\vspace{2mm}
= BR Cir&&(5)&(5)&&(18)&(22)&(22)&&&40, 41\\
XTE J1701--462&Z-source&8.5$\pm$8.0&-&-&-&9$\pm$4     &9$\pm$5&-&0.5&37, 42\\
\vspace{2mm}
              &&(6)        &&&&(23)$^{\ast}$&(6)    &&   &\\
GX 349+2&Z-source&9.25$\pm$0.75&-&-&-&5$\pm$1&7.7$\pm$1.0&-&1.0&7, 37,\\
\vspace{2mm}
= Sco X--2&&(7)&&&&(7)&(31)&&&38\\
XTE J1814--338&MSXP&8.0$\pm$1.6&4.27&-&$\sim 0.5$&0.71$\pm$0.10&1.63$\pm$0.21&-&1.0&19, 37\\
\vspace{2mm}
              &&(8)        &(5)&&(19)&(19)          &(19)          &&&\\
GX 13+1&Z-source&7$\pm$1&-&-&-&15.3$\pm$2.3&-&-&1.0&24\\
\vspace{2mm}
= 4U 1811--17&&(9)&&&&(24)&&&&\\
GX 17+2       &Z-source&8.0$\pm$2.4&-&-&-&12.5$\pm$1.5&15$\pm$2&-&1.0&37, 43\\
\vspace{2mm}
= 4U 1813--14 &&(10)       &&&&(25)        &(32)    &   &&\\
HETE&MSXP&5$\pm$1&1.39&-&0.016--0.07&0.89$\pm$0.22&1.6$\pm$0.4&-&2.0&26, 37,\\
\vspace{2mm}
J1900.1--2455&&(11, 12)&(15)&&(15)&(26)$^{\ast}$&(26)&&&44, 45\\
XTE J2123--058&Atoll&18.4$\pm$2.7&5.96&1.3&0.60&0.37$\pm$0.15&0.66$\pm$0.27&0.77 (R,&-&35, 46\\
\vspace{2mm}
= LZ Aqr&&(5)&(5)&(14)&(14)&(27)&(27)$^{\dagger}$&34, 35)&&\\
Cyg X--2&Z-source&13.4$\pm$2.0&236.2&1.78&0.60&1.24$\pm$0.22&1.9$\pm$0.5&-&1.0&47, 48\\
\vspace{2mm}
= V1341 Cyg&&(5)&(5)&(14)&(14)&(28)&(33)&&&\\
\hline
\end{tabular}
\normalsize
\end{center}
Columns give:
(I) source names;
(II) X-ray classification;
(III) distance estimate;
(IV) orbital period of the system;
(V) mass of the neutron star in solar units (assumed to be $\sim 1.4 M_\odot$ if unconstrained);
(VI) mass of the companion star in solar units;
(VII) and (VIII) interstellar reddening in $V$-band, and interstellar HI absorption column ($^{\ast}A_{\rm V}$ and $^{\dagger}N_{\rm H}$ are estimated here from the relation $N_{\rm H} = 1.79 \times 10^{21} cm^{-2}A_{\rm V}$; \citealt{predet95});
(IX) the companion star OIR luminosity contribution in quiescence;
(X) The maximum time separation, $\Delta t$, between the OIR and X-ray observations defined as quasi-simultaneous;
(XI) References for the quasi-simultaneous OIR and X-ray fluxes collected.
References: 
(1) \cite{torret07};
(2) \cite{boydet00};
(3) \cite{kova00};
(4) \cite*{gallet02};
(5) \cite*{jonket04};
(6) \cite*{kennet06};
(7) \cite*{wachma96};
(8) \cite*{stroet03};
(9) \cite*{bandet99};
(10) \cite*{kuulet02};
(11) \cite*{kawasu05};
(12) \cite*{gallet05b};
(13) \cite{shawet05};
(14) \cite*{rittet03};
(15) \cite{kaaret06};
(16) \cite{gallma05a};
(17) \cite{cramet90};
(18) \cite*{johnet99};
(19) \cite*{krauet05};
(20) \cite{bonnet89};
(21) \cite*{gileet05};
(22) \cite*{jonket07};
(23) \cite*{prodet06};
(24) \cite*{charna92};
(25) \cite*{deutet99};
(26) \cite*{steeet05b};
(27) \cite*{hyneet01};
(28) \cite*{mcclet84};
(29) \cite{schu99};
(30) \cite*{juetet03};
(31) \cite*{iariet04};
(32) \cite*{vrtiet91};
(33) \cite*{costet05};
(34) \cite*{casaet02};
(35) \cite*{shahet03};
(36) \cite{steeet04};
(37) \emph{RXTE} ASM;
(38) this paper;
(39) \cite*{shir98};
(40) \cite*{glas78};
(41) \cite*{mone92};
(42) \cite*{maitba06};
(43) \cite*{callet02};
(44) \cite*{fox05};
(45) \cite*{steeet05a};
(46) \cite*{tomset04};
(47) \cite{hasiet90};
(48) \cite{vanpet90};
(49) \cite*{vanpet80};
(50) \cite*{corbet98};
(51) \cite*{wach97};
(52) \cite*{gilfet98};
(53) \cite*{campet04};
(54) \cite{greeet06};
(55) \cite*{bailet06}
\end{table*}

\subsection{Observations with the Danish 1.54-m Telescope}

$VRIZ$-band imaging of two Z-sources, GX 349+2 and LMC X--2, were taken in July 2001 using the camera on the Danish Faint Object Spectrograph and Camera (\small DFOSC). \normalsize Table 3 lists the observations used in this work. De-biasing and flat fielding was performed with \small IRAF \normalsize and aperture photometry of the targets and the standard star LTT 7987 ($V$ = 12.23 mag; $R$ = 12.29; $I$ = 12.37; $Z$ = 12.50; \citealt{hamuet92}; \citeyear{hamuet94}) was achieved using \small PHOTOM\normalsize. GX 349+2 was detected with a significance of $>8\sigma$ in all exposures (most were $>50\sigma$ detections). For LMC X--2 we aligned and combined the six exposures in each filter on MJD 52115 to achieve a higher signal-to-noise ratio (S/N). No combined images were created from the observations of this source on MJD 52117 as the S/N was sufficiently high for photometry. LMC X--2 was detected at a significance level of $>4\sigma$ in the $Z$-band images and $>20\sigma$ in all other images used in this work. In Fig. 1 we present the finder charts for LMC X--2.

The measured fluxes were accounted for airmass-dependent atmospheric extinction according to \cite{burket95}. We flux-calibrated the data using LTT 7987. For GX 349+2 we measured the flux of three stars in the field of view with known $V$ and $R$-band magnitudes \citep[stars 2, 5 and 7 listed in][]{wachma96} and found their magnitudes to differ from those previously reported by $\sim 0.04$ mag. The resulting fluxes obtained for GX 349+2 and LMC X--2 were de-reddened to account for interstellar extinction using the values of $A_{\rm V}$ listed in Table 1 ($A_{\rm R}$, $A_{\rm I}$ and $A_{\rm Z}$ were calculated according to the recipe in \citealt{paper1}).

\subsection{UKIRT Observations of 4U 0614+09}

NIR imaging of 4U 0614+09 was obtained with the 3.8 m United Kingdom Infrared Telescope (UKIRT) on 2002 February 14 (MJD 52319.3), using UFTI, the UKIRT Fast Track Imager \citep*{rochet03}. Jittered observations of 4U 0614+09 were made in the $J$, $H$ and $K$ filters, with 9$\times$30 s exposures in both $J$ and $H$, and 9$\times$60 s exposures in $K$. The infrared standard star FS 120 ($J$ = 11.335 mag; $H$ = 10.852; $K$ = 10.612) was also observed for a total of 50 s, 25 s and 25 s in $J$, $H$ and $K$, respectively. The airmass of the standard and 4U 0614+09 were very similar: 1.004--1.036.  The `JITTER-SELF-FLAT' data reduction recipe was used, which created a flat field from the sequence of 9 jittered object frames and a dark frame. After dark subtraction and flat fielding, a mosaic was generated from the 9 object frames.

Photometry was carried out using \emph{IRAF}. 4U 0614+09 was detected with a significance of 7.1$\sigma$ in $J$, 15.3$\sigma$ in $H$ and 11.5$\sigma$ in $K$. Flux calibration was achieved using FS 120, yielding the following de-reddened ($A_{\rm V}=1.41$) flux densities for 4U 0614+09: $F_{\rm \nu,J}=0.145 \pm 0.037$ mJy; $F_{\rm \nu,H}=0.139 \pm 0.020$ mJy; $F_{\rm \nu,K}=0.111 \pm 0.022$ mJy (the apparent reddened magnitudes are $J=18.12$; $H=17.50$; $K=16.38$).

\begin{table}
\caption{References of the new data from sources in \citealt{paper1}.}
\begin{tabular}{lllllllll}
\hline
Source&Classifi-&$\Delta t$ /&Fluxes - new   \\
      &cation   &days        &data references\\
\hline
IGR J00291+5934 &MSXP&1.0&1\\
4U 0614+09      &Atoll&0.5&38\\
CXOU 132619.7--&unknown&-&-\\
\multicolumn{1}{c}{472910.8} &&&\\
Cen X--4 &Atoll&0.5&49\\
4U 1608--52 &Atoll&1.0&50, 51\\
Sco X--1 &Z-source&1.0&-\\
SAX J1808.4--3658&MSXP&1.0&37, 52, 53, 54\\
Aql X--1 &Atoll&0.5&37, 55\\
\hline
\end{tabular}
\normalsize
References: see caption of Table 1.
\end{table}

\subsection{Results}

In the upper panels of Figs. 2 and 3 we plot the quasi-simultaneous OIR and X-ray data. The OIR monochromatic luminosity, $L_{\rm \nu, OIR}$ is plotted against $L_{\rm X}^{1/2}a$ (where $a$ is the orbital separation) in Fig. 2 in order to test the X-ray reprocessing model. In Fig. 3, $L_{\rm \nu, OIR}$ is plotted against $L_{\rm X}$: the plot necessary to test the models of jet and viscous disc OIR emission \citep{paper1}. These plots are the same as the left and right panels of Fig. 5 in \citealt{paper1}, with the new NSXB data added and the data from BHXBs removed.

The orbital separations, as with \citealt{paper1}, are inferred from the best known estimates of the orbital period and masses of the neutron star and companion (listed in Table 1, and Table 2 of \citealt{paper1}). For systems with observationally unconstrained neutron star masses, we assume $M_{\rm co}\approx 1.4M_\odot$. We do not include data in Fig. 2 from systems in which the orbital period or the companion mass is unconstrained. We note that data from Cir X--1 in Fig. 2 (open circles in the top right corner of the figure) may not be representative because its orbit is eccentric and so the orbital separation cannot be accurately inferred using this method.

\begin{table*}
\begin{center}
\caption{Log of the Danish 1.54-m Telescope observations.}
\begin{tabular}{lllrrrrrrrr}
\hline
MJD&Target&Exposures&\multicolumn{4}{c}{Integration time / exp.}&\multicolumn{4}{c}{Apparent magnitudes (not de-reddened)}\\
   &      &/filter  &\multicolumn{1}{c}{V}&\multicolumn{1}{c}{R}&\multicolumn{1}{c}{I}&\multicolumn{1}{c}{Z}&\multicolumn{1}{c}{V}&\multicolumn{1}{c}{R}&\multicolumn{1}{c}{I}&\multicolumn{1}{c}{Z}\\
\hline
52114.07--.25 &GX 349+2&9&300&200&120&120&18.46--18.62&17.44--17.62&16.54--16.65&16.04--16.24\\
52114.99--5.07&GX 349+2&4&300&200&120&120&18.32--18.57&17.35--17.53&16.47--16.58&16.07--16.17\\
52115.30--.31 &LTT 7987&1&2  &2  &4  &6  &&&&\\
52115.35--.43 &LMC X--2&6&120&120&180&240&19.19$\pm$0.07&19.01$\pm$0.07&18.91$\pm$0.07&19.12$\pm$0.14\\
52117.37--.44 &LMC X--2&2&120&120&180&300&18.47$\pm$0.07&18.43$\pm$0.07&18.45$\pm$0.08&18.57$\pm$0.26\\
\hline
\end{tabular}
\normalsize
\end{center}
MJD 52114.0 corresponds to 2001-07-24.0 UT. The filters used were Bessel $V$, Bessel $R$, Gunn $i$ and Gunn $z$. LTT 7987 is the standard star used for flux calibration. For GX 349+2 the range of magnitudes measured are tabulated, whereas for LMC X--2 we tabulate the magnitudes from the combined images.
\end{table*}

\begin{figure}
\centering
\includegraphics[width=6cm,angle=0]{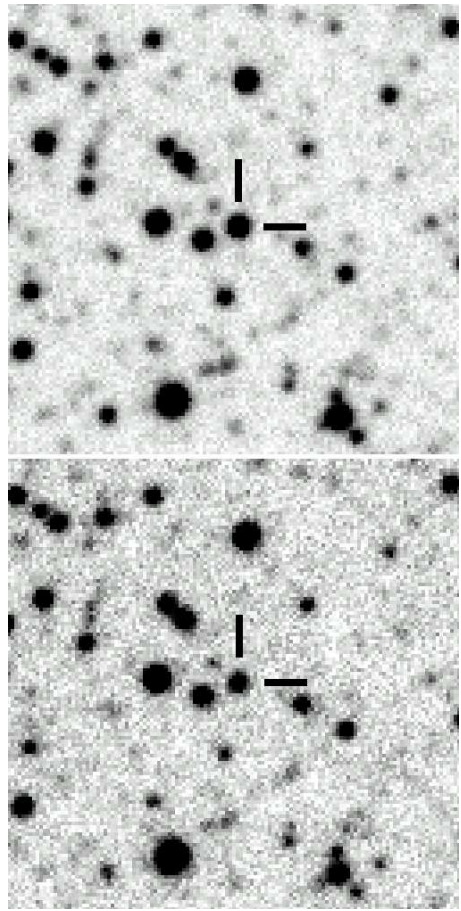}
\caption{High resolution optical finding chart for LMC X--2 in $V$-band (upper panel) and $I$-band (lower panel) \citep[for a lower resolution $B$-band finding chart see][]{johnet79}. North is to the top and east is to the left. The images are $\sim 0.8\times 0.8$ arcmins and were taken on 2001-07-27 with DFOSC on the Danish 1.54-m Telescope.}
\end{figure}

The power law lines in Figs. 2 and 3 represent the expected correlations for X-ray reprocessing \citep{vanpet94} and jet emission \citep[solid line;][]{miglfe06,paper1}, respectively. The jet model is normalised to the observed radio--X-ray data \citep{miglfe06} assuming a flat (optically thick) radio--OIR jet spectrum and the X-ray reprocessing model is normalised to the the optical ($BVRI$) NSXB data of \citealt{paper1}. The dotted power laws in Fig. 2 represent the expected relation for NIR data dominated by X-ray reprocessing assuming an OIR spectral index of $\alpha = 0.5$ and $\alpha = 2.0$ (where $L_{\nu}\propto \nu^{\alpha}$). Although the different orbital inclinations between these sources could change the X-ray and OIR luminosities observed and hence these X-ray--OIR relations, we showed in \citealt{paper1} that inclination doesn't appear to play a significant role.

\begin{figure}
\centering
\includegraphics[width=8.9cm,angle=0]{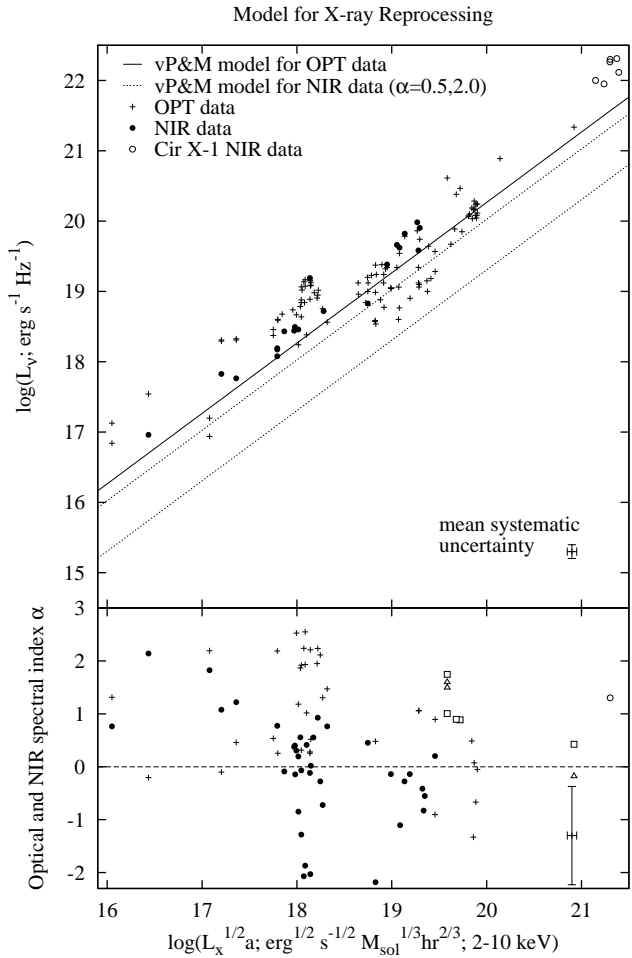}
\caption{Quasi-simultaneous $L_{\rm X}^{1/2}a$ (where $a$ is the orbital separation) versus OIR monochromatic luminosity (upper panel) and optical and NIR spectral index (lower panel) for NSXBs. OPT and NIR refer to the optical and NIR wavebands, respectively (also applies to Fig. 3). In the upper panel, the model for optical emission from X-ray reprocessing (solid line) is derived from vP\&M \citep{vanpet94} and normalised to the optical data of Paper I. The dotted lines show the expected relation for X-ray reprocessing from NIR data, assuming a NIR--optical spectral index of $\alpha = 0.5$ (upper dotted line) and $\alpha = 2.0$ (lower dotted line). In the lower panel, the Z-sources have different symbols to the atolls and MSXPs (also applies to Fig. 3). Data were not included for sources with unknown orbital periods or companion masses (Table 1, and Table 2 of Paper I). Data for Cir X--1 are denoted by open circles; the orbital separation may not be accurate for this source (see Section 2.3).
}
\end{figure}

\begin{figure}
\centering
\includegraphics[width=8.9cm,angle=0]{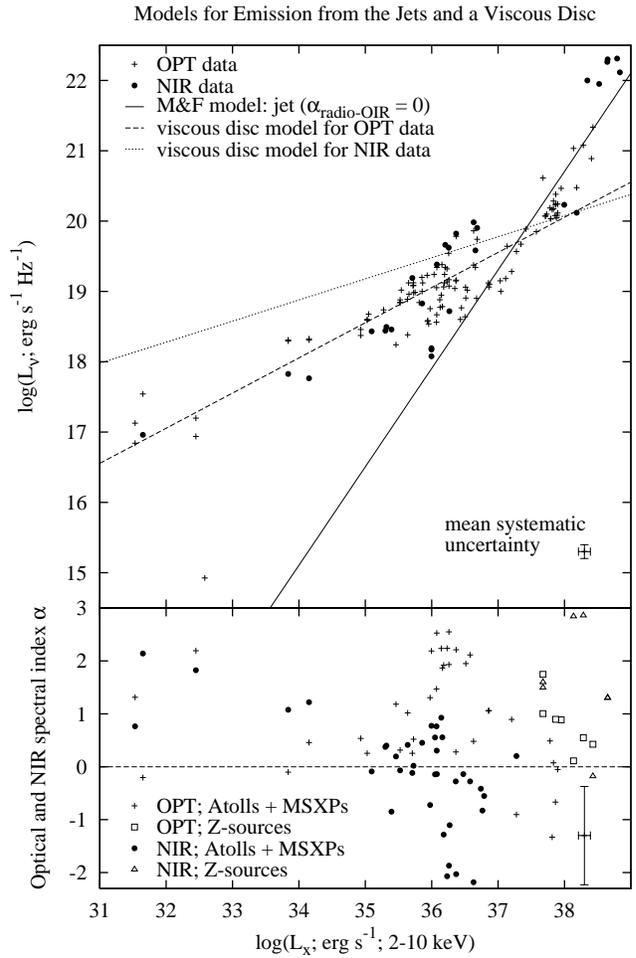}
\caption{Quasi-simultaneous X-ray luminosity versus OIR monochromatic luminosity (upper panel) and optical and NIR spectral index (lower panel) for NSXBs. In the upper panel, the jet model for NSXBs assumes a flat spectrum from radio to OIR and adopts the relation $L_{\rm radio}\propto L_{\rm X}^{1.4}$ \citep[M\&F:][]{miglfe06}. The models for optical and NIR emission from a viscously heated disc are also shown, normalised to the optical and NIR data, respectively.
}
\end{figure}

The lower panels of Figs. 2 and 3 describe the dependence of the shape of the OIR spectrum with $L_{\rm X}^{1/2}a$ and $L_{\rm X}$, respectively. The spectral index $\alpha$ is calculated for all OIR data where two or more optical or NIR data points are quasi-simultaneous. The spectral index of the optical (using data in the $B$, $V$ or $R$-bands) and NIR ($R$, $I$, $J$, $H$ or $K$-bands) data are shown separately. We define the $R$-band as the break between optical and NIR here because it is often the $R-I$ colours that indicate the NIR excess (and for many of these observations $JHK$ data were not obtained). For consistency (and to separate the two wavebands used to calculate $\alpha$) we use the two lowest frequency NIR bandpasses in each OIR SED to calculate $\alpha_{\rm NIR}$ and the two highest frequency optical bands to calculate $\alpha_{\rm OPT}$. Data from the Z-sources are shown as different symbols to the atolls/MSXPs because the spectral index of the former group appears to behave differently to that of the latter (see Section 3.2).

We have also compiled luminosity-calibrated OIR SEDs of 17 NSXBs (Fig. 4). The references for the data are listed in Tables 1 and 2, and Table 2 of \citealt{paper1}. In Section 3 we attempt to interpret the OIR--X-ray plots and OIR SEDs in terms of the dominating optical and NIR emission processes.

\section{Interpretation \& Discussion}

\subsection{OIR -- X-ray Relations}

In the upper panels of Figs. 2 and 3 we improve on Fig. 5 of \citealt{paper1} with a larger sample of NSXB data. Focusing on the upper panel of Fig. 3, we see that the global relation $L_{\rm OIR}\propto L_{\rm X}^{0.6}$ for NSXBs \citep{paper1} does not hold at high X-ray luminosity ($L_{\rm X}\sog 10^{37.5}$ erg s$^{-1}$); the OIR is more luminous than expected from this relation. This is inconsistent with the model for emission from a viscously heated disc \citep{paper1}, which requires a relation $L_{\rm OIR}\propto L_{\rm X}^{0.5}$ for optical data and $L_{\rm OIR}\propto L_{\rm X}^{0.3}$ for NIR (Fig. 3). We can therefore rule out a viscous disc origin to the OIR emission, at least at $L_{\rm X}\sog 10^{37.5}$ erg s$^{-1}$.

The upper panel in Fig. 3 shows that the jet model lies close to the data at high $L_{\rm X}$, suggesting the OIR enhancement at these luminosities may be caused by the domination of the jet. The jet model cannot describe the data at low $L_{\rm X}$ ($\sol 10^{36}$ erg s$^{-1}$). However, the upper panel of Fig. 2 indicates that all the optical data also lie close to the expected relation for X-ray reprocessing in the disc. Hence, only the data above $L_{\rm X}\sog 10^{36}$ erg s$^{-1}$ could arise due to the presence of a jet.

The Z-sources, which tend to have much longer orbital periods (and hence larger orbital separations) than atolls and MSXPs, dominate the highest X-ray luminosities. These sources spend most of their time in a soft X-ray state and in fact have radio luminosities lower than predicted by the NSXB hard state radio--X-ray relation if the radio emission originates in the jet \citep{miglfe06}. We would therefore expect the OIR emission from the jets in Z-sources to also be lower than the model in Fig. 3 unless the radio--OIR jet spectrum is inverted (positive). In fact from Fig. 3 of \cite{miglfe06} we see that at $L_{\rm X} \sim 10^{38}$ erg s$^{-1}$, $L_{\rm radio} \sim 10^{30}$ erg s$^{-1}$ and so from Fig. 3 here, the radio--OIR spectral index for Z-sources\footnote{At $L_{\rm X} \sim 10^{38}$ erg s$^{-1}$ we can calculate the spectral index between radio and OIR for Z-sources since we know the radio and OIR luminosities.} is $\alpha \sim 0.2$. With this information alone, the jets can only dominate the OIR of Z-sources if the radio--OIR spectrum is inverted.

\subsection{OIR Spectral Index -- X-ray Relations}

Since the dominating OIR emission processes in NSXBs are unclear (at least at high $L_{\rm X}$) from the OIR--X-ray relations, we now turn to the OIR spectra. We see from the lower panel of Fig. 3 that there is a relation between $L_{\rm X}$ and $\alpha_{\rm NIR}$ for MSXPs and atolls: the NIR spectrum becomes redder at higher luminosities. Quantitatively, $\alpha_{\rm NIR}$ becomes negative when $L_{\rm X}\sog 10^{36}$ erg s$^{-1}$ and $\alpha_{\rm NIR}$ is positive for all data below $L_{\rm X}\approx 10^{35}$ erg s$^{-1}$. This is opposite to the behaviour of BHXBs, where there is evidence in some systems for the OIR spectrum in the hard state to become redder at low luminosities \citep{paper1}. If the origin of the emission is the disc blackbody, we would expect a bluer (hotter) spectrum at higher luminosities.

The only process expected to produce an OIR spectrum of index $\alpha < 0$ at high luminosities in these systems is optically thin synchrotron. It is therefore intriguing that $\alpha_{\rm NIR} < 0$ for atolls and MSXPs when $L_{\rm X}\sog 10^{36}$ ergs$^{-1}$ s$^{-1}$; the X-ray luminosity range in which the jet could play a role (Fig. 3, upper panel). Since there are just five NIR data points in the lower panel of Fig. 3 below $L_{\rm X} = 10^{35}$ erg s$^{-1}$, we perform a Kolmogorov-Smirnov (K-S) test to quantify the significance of the apparent $\alpha_{\rm NIR}$--$L_{\rm X}$ relation for the atolls/MSXPs. We use the `Numerical recipes in \small FORTRAN\normalsize ' \citep{preset92} routine `kstwo' which is the K-S test for two data sets, to determine if the values of $\alpha_{\rm NIR}$ differ significantly below and above $L_{\rm X} = 10^{35}$ erg s$^{-1}$. The maximum difference between the cumulative distributions is $D = 0.91$ with a corresponding probability of $P = 5.0\times 10^{-4}$; i.e. the probability that the NIR spectral index of the data below $L_{\rm X} = 10^{35}$ erg s$^{-1}$ belongs to the same population as the data above $L_{\rm X} = 10^{35}$ erg s$^{-1}$ is 0.05 percent.

In addition, we have carried out a Spearman's Rank correlation test on the $\alpha_{\rm NIR}$--$log~L_{\rm X}$ and $\alpha_{\rm OPT}$--$log~L_{\rm X}$ data of atolls/MSXPs (Table 4). We find an anti-correlation between $\alpha_{\rm NIR}$ and $log~L_{\rm X}$ at the 3.8$\sigma$ confidence level, supporting the above K-S test results. We do not find a significant relation between $\alpha_{\rm OPT}$ and $log~L_{\rm X}$ except when we impose a somewhat arbitrary X-ray luminosity cut: for data above $L_{\rm X}> 10^{36}$ erg s$^{-1}$ there is an anti-correlation at the 3.5$\sigma$ confidence level. This again could be due to the jet contribution dominating at these highest luminosities. The confidence of this result should be taken with caution as it could be dominated by the group of data with the highest $\alpha_{\rm OPT}$ values which happen to lie just above the $L_{\rm X}= 10^{36}$ erg s$^{-1}$ cut.

\begin{table}
\caption{Results of the Spearman's Rank correlation between $\alpha$ and $log~L_{\rm X}$ for atolls/MSXPs.}
\begin{tabular}{llll}
\hline
OIR data&Range in $L_{\rm X}$&Correlation&Significance\\
used    &(erg s$^{-1}$)      &coefficient&            \\
\hline
$\alpha_{\rm NIR}$&all                 &$r_{\rm s}=-0.63$&3.8$\sigma$\\
$\alpha_{\rm OPT}$&all                 &$r_{\rm s}=-0.14$&0.8$\sigma$\\
$\alpha_{\rm OPT}$&$L_{\rm X}> 10^{36}$&$r_{\rm s}=-0.77$&3.5$\sigma$\\
\hline
\end{tabular}
\normalsize

\end{table}

The optical spectral index $\alpha_{\rm OPT}$ is generally positive for atolls and MSXPs (Fig. 3 lower panel), but decreases at $L_{\rm X} \sog 10^{37}$ erg s$^{-1}$. We would expect $\alpha_{\rm OPT}$ to become negative at a higher X-ray luminosity than $\alpha_{\rm NIR}$ if the origin of the redder emission component is the jets. As the X-ray luminosity is increased, the optically thin synchrotron jet component will dominate over X-ray reprocessing in the NIR bands before the optical as the jet component has a negative spectral index. At $L_{\rm X} > 10^{37}$ erg s$^{-1}$, $\alpha_{\rm OPT}$ is negative in 63 percent of the optical data of atolls/MSXPs. Below $L_{\rm X} = 10^{37}$ erg s$^{-1}$ this is 7 percent. These few data in the latter group with $\alpha_{\rm OPT} < 0$ are at low $L_{\rm X}$ ($< 10^{34}$ erg s$^{-1}$) and may be due to cooler accretion discs \citep[possibly like the BHXBs;][]{paper1}. However, the mean uncertainty in the values of $\alpha$ are fairly large, so we can only make conclusions from general trends and not individual data points.

\begin{figure*}
\includegraphics[height=20cm,angle=0]{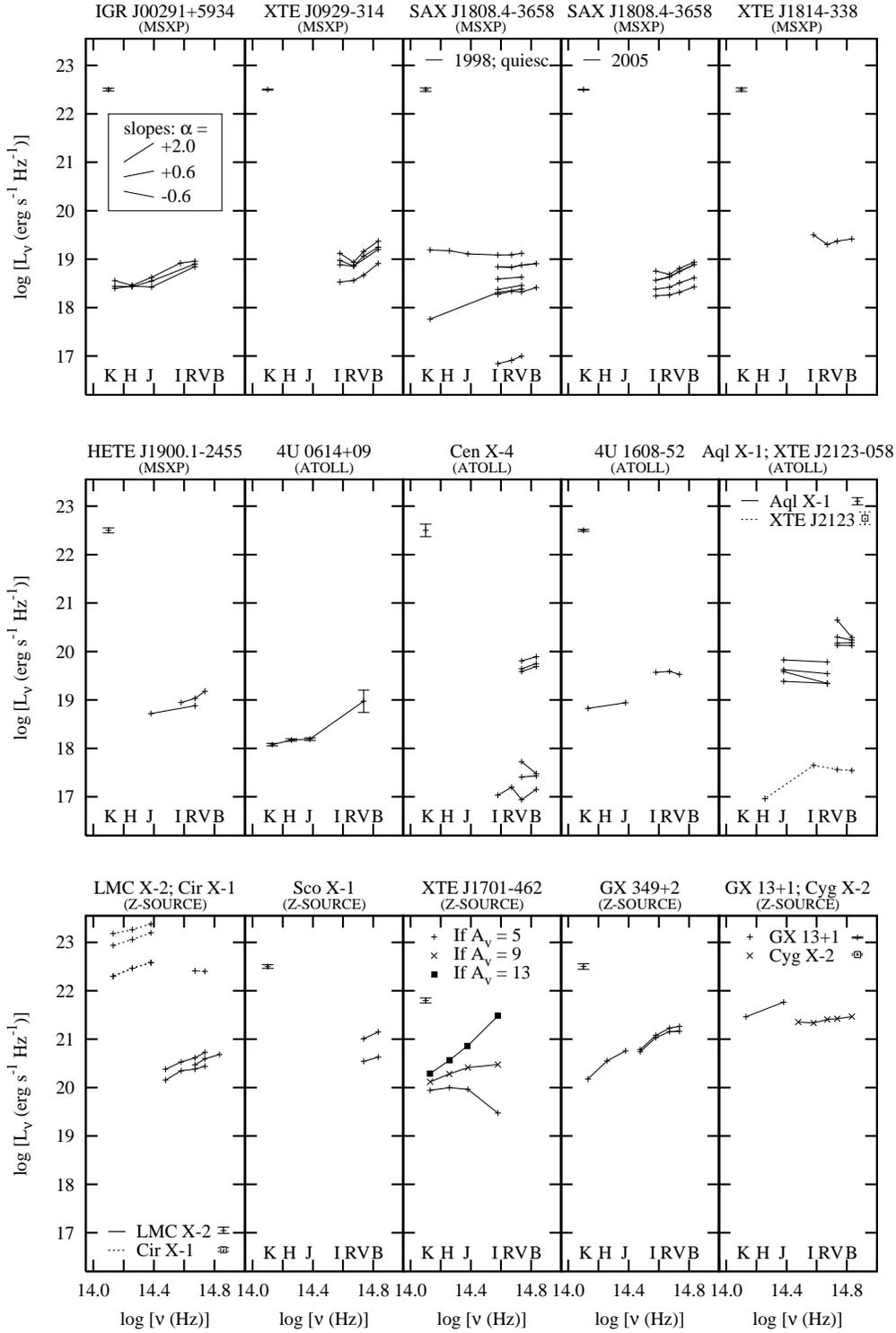}
\caption{Spectral Energy Distributions (SEDs) of 17 NSXBs. The key in the top left panel corresponds to the slope of the continuum, i.e. the spectral index $\alpha$ (where $L_{\nu} \propto \nu^{\alpha}$). MSXP, ATOLL and Z-SOURCE refer to millisecond X-ray pulsars, atoll sources and Z-sources, respectively. The references for the data are listed in Tables 1 and 2 and Table 2 of \citealt{paper1}. The mean systematic error for the data of each source is indicated in each panel (in the top left corner in most cases).
}
\end{figure*}

Almost all of the spectra of the Z-sources are blue ($\alpha > 0$; Fig. 3 lower panel). This supports the suggestion in Section 3.1 that X-ray reprocessing dominates the OIR of the Z-sources due to their larger accretion discs and lower radio jet luminosities (at a given $L_{\rm X}$). The Z-sources cannot be dominated by optically thin synchrotron emission as this requires $\alpha < 0$. Since the radio--OIR spectrum of Z-sources is $\alpha \approx 0.2$ (Section 3.1), the optically thick part of the jet spectrum could dominate the OIR if $\alpha_{\rm OIR} \approx 0.2$ also (although this requires the optically thick/thin break to be in the optical regime or bluer). $\alpha_{\rm OPT,NIR} > 0.2$ is observed for most of the data from the Z-sources (Fig. 3 lower panel), which implies this is not the case, however we cannot rule out an optically thick jet origin to a few of the OIR Z-source data (those with lower spectral index).

The lower panel of Fig. 2 shows how the optical and NIR spectral index changes with $L_{\rm X}^{1/2}a$. The Z-source data in this panel have the largest values of $L_{\rm X}^{1/2}a$. The expected level of OIR emission from reprocessing in the disc in these systems is larger than that of the atolls/MSXPs due to their larger accretion discs and higher X-ray luminosities, further supporting the X-ray reprocessing scenario.

\subsection{The OIR SEDs}

The SEDs of individual sources in Fig. 4 allow us to see how the luminosity-dependent optical and NIR spectral indices differ between sources. We group the panels in Fig. 4 into the different types of NSXB: MSXPs in the upper panels, then atolls, then Z-sources in the lower panels.

One can visually see a NIR excess ($\alpha < 0$) joining a blue ($\alpha > 0$) optical spectrum in the SEDs of four of the five MSXPs (the `V'-shape). This is indicative of two separate spectral components and is observed in at least three hard state BHXBs \citep{corbet02,homaet05,paper1}, where it is interpreted as the optically thin jet spectrum meeting the thermal spectrum of the accretion disc. The NIR excess disappears at low luminosities in a total of four outbursts of three MSXPs; no NIR excess is seen below $L_{\rm \nu,OIR} < 10^{18}$ erg s$^{-1}$ Hz$^{-1}$ in any MSXP or in fact any NSXB.

Of the five atolls in Fig. 4, one (Aql X--1; the atoll with the most data and highest luminosity) has a negative spectral index; more-so at high luminosity. There is little NIR coverage of atolls in the literature (at least at high luminosities) and from the data we have, the OIR SEDs are positive in all sources except Aql X--1, consistent with the disc dominating the OIR. The optical spectral index for Aql X--1 is very negative at the highest luminosities, suggesting all of the OIR was dominated by the jets during the peak of the bright 1978 outburst of this source. The SED of 4U 0614+09 flattens (becomes redder) in the NIR; this is known to be where the disc spectrum meets that of the jet \citep{miglet06}.

There is no evidence for a jet ($\alpha < 0$) component contributing to the OIR spectra of any of the seven Z-sources in Fig. 4. As was shown above, the jets should be OIR-bright in Z-sources but because of their larger discs, X-ray reprocessing dominates. For the new transient Z-source XTE J1701--462 \citep{homaet07}, the optical extinction is not well constrained and we show three SEDs in Fig. 4 of the same data, using different values of $A_{\rm V}$ to illustrate the possible range in spectral indices (we have not included any data from XTE J1701--462 in the lower panels of Figs. 2 or 3 because of these uncertainties in $\alpha$).

\subsection{The Full Picture}

From the information gathered in Sections 3.1--3.3, the picture is now emerging of the dominating OIR emission processes for the different types of NSXB at different luminosities. For NSXBs at low luminosities ($L_{\rm X} \sol 10^{36}$ erg s$^{-1}$), the correlations and spectra are consistent with X-ray reprocessing dominating the OIR. The data are inconsistent with emission from the jets at these luminosities.

We can rule out a viscously heated disc origin to the NIR data of all NSXBs since that requires $L_{\rm NIR} \propto L_{\rm X}^{\sim 0.3}$ which is not observed (Fig. 3). Similarly the viscous disc cannot dominate the optical light of NSXBs at $L_{\rm X} \sog 10^{37.5}$ erg s$^{-1}$ but could below this luminosity as $L_{\rm OPT} \propto L_{\rm X}^{\sim 0.5}$ is required and observed. The jets dominate the NIR and optical light of atolls/MSXPs above $L_{\rm X} \approx 10^{36}$ and $L_{\rm X} \approx 10^{37}$ erg s$^{-1}$ respectively, as the spectral indices of the continuum cannot be explained by thermal emission. The SEDs of the transient NSXBs with the best sampling show the physical disappearance at low luminosity of a NIR excess which is present at high luminosity. These observations are also consistent with the jet OIR--X-ray model.

For Z-sources, the OIR--X-ray relations and OIR spectra are consistent with emission from X-ray reprocessing. The jets and viscously heated disc can be ruled out in most Z-sources, however the optical and NIR spectral indices in a few observations are also consistent with an optically thick jet which extends from the (measured) radio flux.

We can make a direct measurement of the jet radio--NIR spectral index for the atolls/MSXPs, using the NIR data which are dominated by the jets. In the luminosity range $10^{36} < L_{\rm X} < 10^{37}$ erg s$^{-1}$, the NIR data with a negative spectral index is on average 0.70 dex more luminous than expected from the jet model \citep[which assumes a flat radio--NIR spectrum; we know the radio luminosity at this $L_{\rm X}$ from][]{miglfe06}. Some low level contribution from the disc could only partly explain this excess. The corresponding radio--NIR spectral index in that range of $L_{\rm X}$ is $\alpha \approx 0.16$ (0.70 dex in luminosity divided by 4.5 dex in frequency between radio and NIR). The optically thick to optically thin break frequency in NSXBs is thought to be in the mid-IR \citep{miglet06}, making the optically thick radio--mid-IR jet spectrum more inverted; $\alpha_{\rm thick} \geq 0.2$. The spectral evidence for the existence of synchrotron-emitting jets in NSXBs presented in this paper \citep[and showed for the first time by][]{miglet06}, along with radio detections which are sometimes resolved \citep[see][]{fomaet01,miglfe06}, have direct implications for the local conditions and accretion processes in NSXBs. For example, according to \cite{mass06}, a jet can only occur if the magnetic field intensity and mass accretion rate are constrained thus:
\begin{eqnarray*}
  0.87 \left( \frac{B}{10^8 G}\right)^{4/7} \left( \frac{\dot{m}}{10^{-8} M_\odot yr^{-1}}\right)^{-2/7} \leq 1\\
  \end{eqnarray*}

From Table 1 of \cite{mass06}, MSXPs only satisfy this condition if we adopt the largest mass accretion rates and smallest magnetic field strengths measured. Empirically we do see the jet only at high luminosities and hence high mass accretion rates. The atolls and Z-sources more easily satisfy the condition for jet production, except for the highest magnetic field strengths measured in Z-sources. Since we know jets exist in all three flavours of NSXB at least at high luminosities, perhaps the measured mass accretion rates and magnetic field strengths are under- and over-estimated respectively, in some cases.

\section{Conclusions}

We have shown that the dominating OIR emission processes in NSXBs vary with X-ray luminosity and between source types (atolls/MSXPs and Z-sources). However, a clearer picture seems to emerge than the dominating emission processes of BHXBs \citep{paper1}. Models predict that X-ray reprocessing in the accretion disc should dominate the OIR at low luminosities and the jets, if present, should dominate at high luminosities for NSXBs with relatively small accretion discs.

From the observed spectral index of the OIR continuum and from OIR--X-ray relations we show that this is the case in atolls and MSXPs: the jets dominate the NIR and optical emission above $L_{\rm X} \approx 10^{36}$ erg s$^{-1}$ and $L_{\rm X} \approx 10^{37}$ erg s$^{-1}$, respectively. Below these luminosities X-ray reprocessing dominates, although we cannot rule out a viscously heated disc origin to the optical emission. We have shown that the radio--NIR spectral index of the jets in the atolls/MSXPs is slightly inverted: $\alpha \approx 0.16$, at least at high X-ray luminosities ($10^{36}$--$10^{37}$ erg s$^{-1}$). In the Z-sources, which have larger discs, we find that X-ray reprocessing is responsible for all OIR emission, and the radio--OIR jet spectrum has to be $\alpha \leq 0.2$ (otherwise the jet spectrum would dominate over the disc). However, the optically thick part of the jet spectrum could dominate in a few cases.

Evidence for the existence of NSXB jets is mounting both from radio detections \citep[which are sometimes resolved; see e.g.][]{fomaet01} and spectrally from higher frequency observations (Migliari et al. 2006; this paper). The power carried in NSXB jets is sensitive to the position of the break between optically thick and thin emission in its spectrum, and is currently a topic of debate. This is constrained in 4U 0614+09 \citep{miglet06}, where it is likely to lie in the mid-infrared. Mid-infrared and NIR photometry and polarimetry could shed light on this; optically thin synchrotron emission is expected to be highly polarised if the magnetic field is ordered.

\vspace{5mm}
\emph{Acknowledgements}.
Based on observations made with the Danish 1.54-m Telescope at the La Silla Observatory and the United Kingdom Infrared Telescope, which is operated by the Joint Astronomy Centre on behalf of the UK Particle Physics and Astronomy Research Council. We thank the anonymous referee for thorough comments on the manuscript.

\end{document}